\def\simlt{\hbox{ \rlap{\raise 0.425ex\hbox{$<$}}\lower 0.65ex\hbox{$\sim$} }}
\def\simgt{\hbox{ \rlap{\raise 0.425ex\hbox{$>$}}\lower 0.65ex\hbox{$\sim$} }}
\def\etal{{\it et al.}}
\def\h{\hskip 1cm}

\documentstyle[11pt,aaspp4,epsf,lscape,longtable]{article}

\lefthead{Nikolaev \etal{}}
\righthead{2MASS Calibration}

\begin{document}

\title{A Global Photometric Analysis of 2MASS Calibration Data}
\author{Sergei Nikolaev\altaffilmark{1}, Martin D. Weinberg\altaffilmark{1}, 
Michael F. Skrutskie\altaffilmark{1}, \\
Roc M. Cutri\altaffilmark{2},
Sherry L. Wheelock\altaffilmark{2}, John E. Gizis\altaffilmark{2}, 
and Eric M. Howard\altaffilmark{1}}
\altaffiltext{1}{Department of Astronomy, University of Massachusetts, 
Amherst MA 01003-4525 \\
E-mail: nikolaev@redtail.astro.umass.edu, weinberg@osprey.astro.umass.edu, 
skrutski@north.astro.umass.edu, ehwd@kutath.astro.umass.edu}
\altaffiltext{2}{Infrared Processing and Analysis Center,
California Institute of Technology, Pasadena, CA 91125 \\
E-mail: roc@ventus.ipac.caltech.edu, slw@castor.ipac.caltech.edu,
gizis@whitesands.ipac.caltech.edu}

\begin{abstract}
We present results from the application of a global photometric calibration 
(GPC) procedure to calibration data from the first 2 years of The Two Micron 
All Sky Survey (2MASS).  The GPC algorithm uses photometry of both primary 
standards and moderately bright `tracer' stars in 35 2MASS calibration fields.
During the first two years of the Survey, each standard was observed on 
approximately 50 nights, with about 900 individual measurements.  Based on 
the photometry of primary standard stars and secondary tracer stars and under 
the assumption that the nightly zeropoint drift is linear, GPC ties together 
all calibration fields and all survey nights simultaneously, producing a globally 
optimized solution.  Calibration solutions for the Northern and Southern 
hemisphere observatories are found separately, and are tested for global 
consistency based on common fields near the celestial equator.

Several results from the GPC are presented, including establishing candidate 
secondary standards, monitoring of near-infrared atmospheric extinction 
coefficients, and verification of global validity of the standards.  The
solution gives long-term averages of the atmospheric extinction coefficients, 
$A_J=0.096$, $A_H=0.026$, $A_{K_s}=0.066$ (North) and $A_J=0.092$, $A_H=0.031$, 
$A_{K_s}=0.065$ (South), with formal error of $0.001$.  The residuals show small 
seasonal variations, most likely due to changing atmospheric content of water 
vapor.  Extension of the GPC to $\sim 100$ field stars in each of the 35 
calibration fields yields a catalog of more than two thousand photometric 
standards ranging from $10^{th}$ to $14^{th}$ magnitude, with photometry that 
is globally consistent to $\sim 1\%$.
\end{abstract}

\keywords{methods: data analysis --- standards --- surveys}


\section{Introduction}

The Two Micron All Sky Survey (2MASS) maps the entire sky in three 
near-infrared (NIR) bands, $J$ ($1.25 \mu$m), $H$ ($1.65 \mu$m), and 
$K_s$\footnote{The $K_s$ band (pronounced `K-short') is described in 
e.g., Persson \etal{} (1998).} ($2.16 \mu$m).  The survey operates in both 
hemispheres, with the Northern facility at Mt. Hopkins, AZ and the Southern 
site at Cerro Tololo, Chile.  Both sites acquire data with nearly identical 
$1.3$ m cassegrain equatorial telescopes 
optimized for efficient sky coverage.  Each telescope is equipped with three
NICMOS3 arrays capable of simultaneous observations in three NIR bands.  The 
Northern 2MASS facility began routine survey observations in June 1997, while
the Southern facility started in March 1998.  

   Nightly photometric calibration for 2MASS is derived using repeated
observations of $1^\circ \times 8.5'$ calibration ``fields'' (see 
Table~\ref{table1}).  Each hour, one of 40 calibration fields is scanned 
six times in the normal survey mode, providing six independent measurements 
of the standard stars in each field.  The 2MASS calibration fields were 
selected initially to contain at least one ``primary'' calibration star drawn 
from the list of faint infrared standards published by Persson \etal{} (1998) 
and Casali \& Hawarden (1992).  One of the fields, 92049, does not contain 
an a priori standard star, but was selected to fill a gap in the right 
ascension coverage of the fields.  Photometry for the calibration stars in 
this field was developed using the techniques described in \S 3.1 of 
this work.  Table~\ref{table1} contains a listing of the 2MASS calibration 
fields and the a priori catalog magnitudes for the primary calibration stars 
in each.  The fields centers listed are the coordinates of the primary 
calibration stars.  The analysis described in this paper uses 2MASS pipeline
source extractions and photometry from repeated observations of 35 of the 
calibration fields.

{\small
\begin{table*}[p]
\begin{centering}
\caption{2MASS calibration fields.  Each field is $8.5'\times1^\circ$, centered
on given coordinates of the fiducial standard.  The equatorial fields 
are observed by both 2MASS facilities in the course of the survey to ensure 
uniformity between hemispheres.  The listed magnitudes are the ones contained
in the literature. \label{table1}}
\begin{tabular}{llcccccc}
\hline\hline
Number & Name & R.A. (J2000.0) & Dec. (J2000.0) & $J$ & $H$ & $K_s$ & $N_{src}$ \\
\hline
\multicolumn{8}{c}{Zenith (North):} \\
90091$^{\rm a}$ & P091-D   & $09:42:58.7$ &$+59:03:43$ & 11.676& 11.348& 11.276& 76 \\
90182$^{\rm a}$ & P182-E   & $18:39:33.8$ &$+49:05:38$ & 12.106& 11.774& 11.701& 253 \\
90161$^{\rm a}$ & P161-D   & $07:00:52.0$ &$+48:29:24$ & 11.695& 11.418& 11.369& 192 \\
90290$^{\rm a}$ & P290-D   & $23:30:33.5$ &$+38:18:57$ & 11.641& 11.362& 11.281& 204 \\
90247$^{\rm a}$ & P247-U   & $03:32:03.0$ &$+37:20:39$ & 11.952& 11.626& 11.530& 394 \\
90272$^{\rm a}$ & P272-D   & $14:58:33.2$ &$+37:08:33$ & 11.632& 11.284& 11.214& 74 \\
90266$^{\rm a}$ & P266-C   & $12:14:25.4$ &$+35:35:56$ & 11.631& 11.370& 11.322& 66 \\
90330$^{\rm a}$ & P330-E   & $16:31:33.6$ &$+30:08:48$ & 11.811& 11.489& 11.430& 126 \\
92409$^{\rm e}$ & Abell2409& $22:00:28.0$ &$+20:51:00$ &  ---  &  ---  &  ---  & 252 \\
\multicolumn{8}{c}{Equatorial:} \\
90067$^{\rm b}$ & M67          & $08:51:14.1$ &$+11:50:52$ & 13.013& 12.724& 12.662& 425 \\
90533$^{\rm a}$ & P533-D       & $03:41:02.4$ &$+06:56:13$ & 11.750& 11.430& 11.353& 111 \\
90565$^{\rm a}$ & P565-C       & $16:26:42.7$ &$+05:52:20$ & 12.171& 11.903& 11.840& 189 \\
90191$^{\rm c}$ & LHS191       & $04:26:20.6$ &$+03:37:25$ & 11.620& 11.038& 10.699& 150 \\
90004$^{\rm b}$ & FS4          & $01:54:37.8$ &$+00:43:02$ & 10.573& 10.328& 10.294& 83 \\
90893$^{\rm a}$ & S893-D       & $23:18:10.1$ &$+00:32:55$ & 11.403& 11.123& 11.063& 86 \\
90013$^{\rm b}$ & FS13         & $05:57:07.6$ &$+00:01:11$ & 10.524& 10.201& 10.149& 374 \\
90860$^{\rm a}$ & S860-D       & $12:21:39.4$ &$-00:07:13$ & 12.190& 11.905& 11.852& 66 \\
90867$^{\rm a}$ & S867-V       & $14:40:58.0$ &$-00:27:47$ & 12.023& 11.681& 11.610& 99 \\
90868$^{\rm c}$ & T868-53850   & $15:00:26.4$ &$-00:39:28$ & 11.595& 10.991& 10.641& 113 \\ 
92026$^{\rm c}$ & LHS2026      & $08:32:30.0$ &$-01:34:14$ & 12.078& 11.485& 11.149& 206 \\
90021$^{\rm c}$ & BRI0021-0214 & $00:24:24.6$ &$-01:58:22$ & 11.864& 11.074& 10.561& 81 \\
90547$^{\rm c,d}$ & L547         & $18:51:17.9$ &$-04:16:28$ & 11.872&  9.831&  8.870& 7302 \\
90808$^{\rm a}$ & S808-C       & $19:01:55.5$ &$-04:29:12$ & 10.936& 10.620& 10.545& 3575 \\
90813$^{\rm a}$ & S813-D       & $20:41:05.2$ &$-05:03:43$ & 11.473& 11.137& 11.077& 276 \\
92202$^{\rm c}$ & BRI2202-1119 & $22:05:35.8$ &$-11:04:29$ & 11.666& 11.077& 10.736& 117 \\
92397$^{\rm c}$ & LHS2397a     & $11:21:49.1$ &$-13:13:13$ & 11.897& 11.190& 10.709& 100 \\
\multicolumn{8}{c}{Zenith (South):} \\
90009$^{\rm c}$ & Oph-n9a  & $16:27:13.2$ &$-24:41:23$ & 15.381& 12.278& 10.733& 244 \\
90312$^{\rm a}$ & S312-T   & $08:25:36.2$ &$-39:05:58$ & 11.949& 11.669& 11.609& 1358 \\
90294$^{\rm a}$ & S294-D   & $00:33:15.3$ &$-39:24:10$ & 10.914& 10.637& 10.582& 65 \\
90301$^{\rm a}$ & S301-D   & $03:26:53.9$ &$-39:50:38$ & 12.153& 11.842& 11.788& 87 \\
90273$^{\rm a}$ & S273-E   & $14:56:52.0$ &$-44:49:14$ & 11.301& 10.896& 10.850& 661 \\
90279$^{\rm a}$ & S279-F   & $17:48:22.7$ &$-45:25:45$ & 12.457& 12.124& 12.034& 1846 \\
90234$^{\rm a}$ & S234-E   & $20:31:20.5$ &$-49:38:59$ & 12.464& 12.127& 12.070& 187 \\
90217$^{\rm a}$ & S217-D   & $12:01:45.2$ &$-50:03:10$ & 11.294& 11.000& 10.923& 595 \\
90121$^{\rm a}$ & S121-E   & $06:29:29.4$ &$-59:39:31$ & 12.114& 11.838& 11.781& 207 \\
\hline
\end{tabular}
\par\parbox[t]{16cm}{
\vskip 0.1cm
$^{\rm a}$NICMOS Standards (Persson \etal{} 1998);

$^{\rm b}$UKIRT Faint Standards (Casali \& Hawarden 1992);

$^{\rm c}$Faint Red Standards (Persson \etal{} 1998);

$^{\rm d}$Based on 2MASS observations, red standard L547 was found variable 
and was subsequently replaced by another star, selected from secondary 
standards, see \S\ref{sec:secondaries}.

$^{\rm e}$Internal calibration field defined by the 2MASS Project; contains no
fiducial standard.}
\end{centering}
\end{table*}
}

The paper has three major parts: \S\ref{sec:method}, which introduces 
the global photometric calibration method and compares it to nightly 
calibration; \S\ref{sec:applications}, which presents the results of 
major applications of the procedure to 2MASS data; and \S\ref{sec:summary}, 
which restates and summarizes the main results of the analysis.  The 
major applications of GPC are: (i) establishment of secondary standards 
(\S\ref{sec:secondaries}), (ii) study of atmospheric extinction in 
near-infrared (\S\ref{sec:extinction}), and (iii) analysis of global 
consistency (\S\ref{sec:uniformity}).  Technical aspects of the method 
are given in Appendix~\ref{sec:details}.


\section{Global Photometric Calibration (GPC)} \label{sec:method}

\begin{figure}
\epsfysize=16.5cm
\centerline{\epsfbox{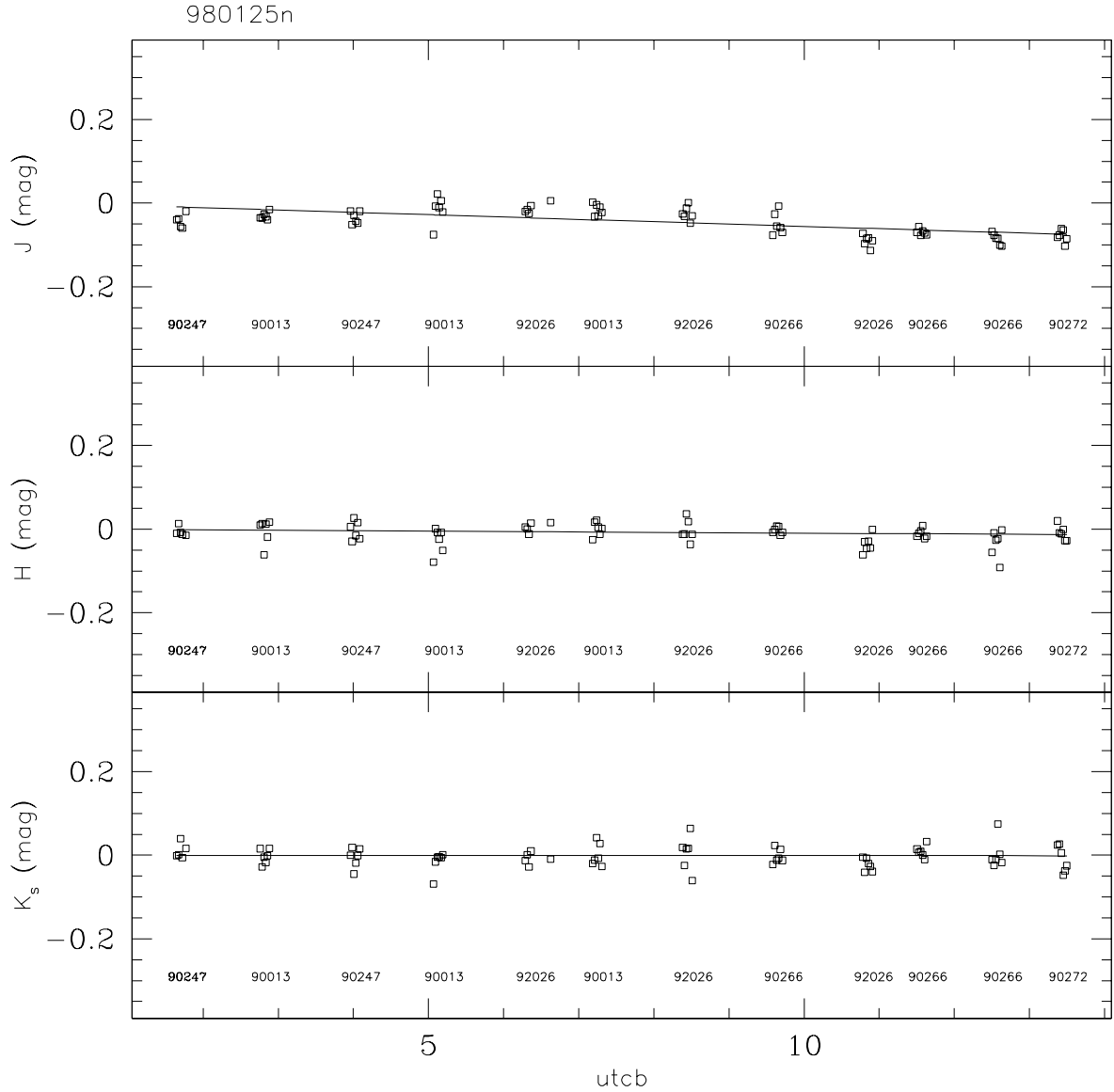}}
\caption{Nightly calibration solutions in three bands for a typical night.
The $x$-axis units are hours.  Data points represent calibrated photometry 
of primary standards, $m_{cal}-m_{inst}+C_2\, (X-1)$, straight lines show 
photometric zero point $C_1$ (see text).  Calibration fields are indicated.  
Note the second order effect in $J$ band (non-linear behavior of sensitivity), 
which is not addressed by the procedure but tends to be small.  
\label{fig:calnight}}
\end{figure}

The purpose of {\em nightly} photometric calibration is to transfer instrumental
magnitudes, $m_{inst}$, onto a uniform photometric system $m_{cal}$:
\begin{equation}
m_{cal} = m_{inst} + C_1 - C_2\, (X-1), \label{eq1}
\end{equation}
where $C_1$ is the photometric zero point, $C_2$ is the atmospheric extinction 
correction, and $X$ is the airmass.  
Figure~\ref{fig:calnight} plots a typical night of calibration data, showing 
both the data points (photometry of primary standards) and the nightly 
calibration solution $C_1$ (straight line).  Each cluster of points in the 
nightly calibration plot is derived from one set of calibration observations.
Each cluster consists of six data points which plot the zeropoint offset (i.e., 
$m_{cal}-m_{inst}+C_2\,(X-1)$) derived for each of the six apparitions of the primary 
calibration star.  Because the Survey's 2.0$^{\prime\prime}$ camera pixels are 
larger than the typical $\sim$1$^{\prime\prime}$ seeing disk, pixelization flux 
errors dominate the scatter of the bright primary standard observations and limit 
the rms uncertainty for multiple apparitions of bright stars in 2MASS observations 
to $\sim 0.02-0.03$ magnitude -- underscoring the advantage of using several flux 
calibrators in each field to minimize the zeropoint uncertainty of a single 
calibration observation.

Experience has shown that the nightly photometric zero point $C_1$ is usually
either a constant, or a linear function of time,  $C_1 (t) = A + B t$.
Equation~(\ref{eq1}) is written for a set of primary calibrators with known
$m_{cal}$ (Table~\ref{table1}), independently for each NIR band.  Solving the
calibration equation~(\ref{eq1}) by a standard least squares algorithm produces
a nightly photometric calibration solution, i.e. $C_1$ and $C_2$.

The main difference between nightly and {\em global} calibration procedures 
is in the
scope of the data upon which the solution is based: the former derives the
calibration parameters $C_1$ and $C_2$ from a single night of data, while
the latter uses all survey nights.  The calibration equation~(\ref{eq1}) is the
same in both cases, but for GPC it can be rewritten to underscore
its global nature:
\begin{equation}
m^{inst}_{ijkn} = m^{cal}_{ik} + a_n + b_n \Delta t_{jn} + A\, (X_{jn}-1), 
\label{eq2}
\end{equation}
where $m^{inst}_{ijkn}$ is the instrumental magnitude of the $i^{th}$ star in 
the $k^{th}$ calibration field, observed at time moment $t_j$ during the $n^{th}$ 
survey night, $m^{cal}_{ik}$ is the true photometric magnitude of the star,
$a_n$ and $b_n$ are nightly photometric constants (one pair for each night),
and $\Delta t_{jn}$ is the time offset between time $t_j$ and midnight.
The atmospheric extinction coefficient is now denoted by~$A$.  
Equation~(\ref{eq2}) implicitly assumes that atmospheric extinction is 
constant.  Seasonal variations in extinction coefficients (see 
\S\ref{sec:extinction}) are thereby smoothed and represented by their 
average values.  The equation is applied to each photometric band separately
and produces one solution for each band.  Note that the global procedure does not 
rely on a priori knowledge of $m_{cal}$: the procedure can solve for these
magnitudes as well.  However, in that case the calibrated magnitudes will
be determined to within an arbitrary constant offset.

\begin{figure}
\epsfysize=8.5cm
\centerline{\epsfbox{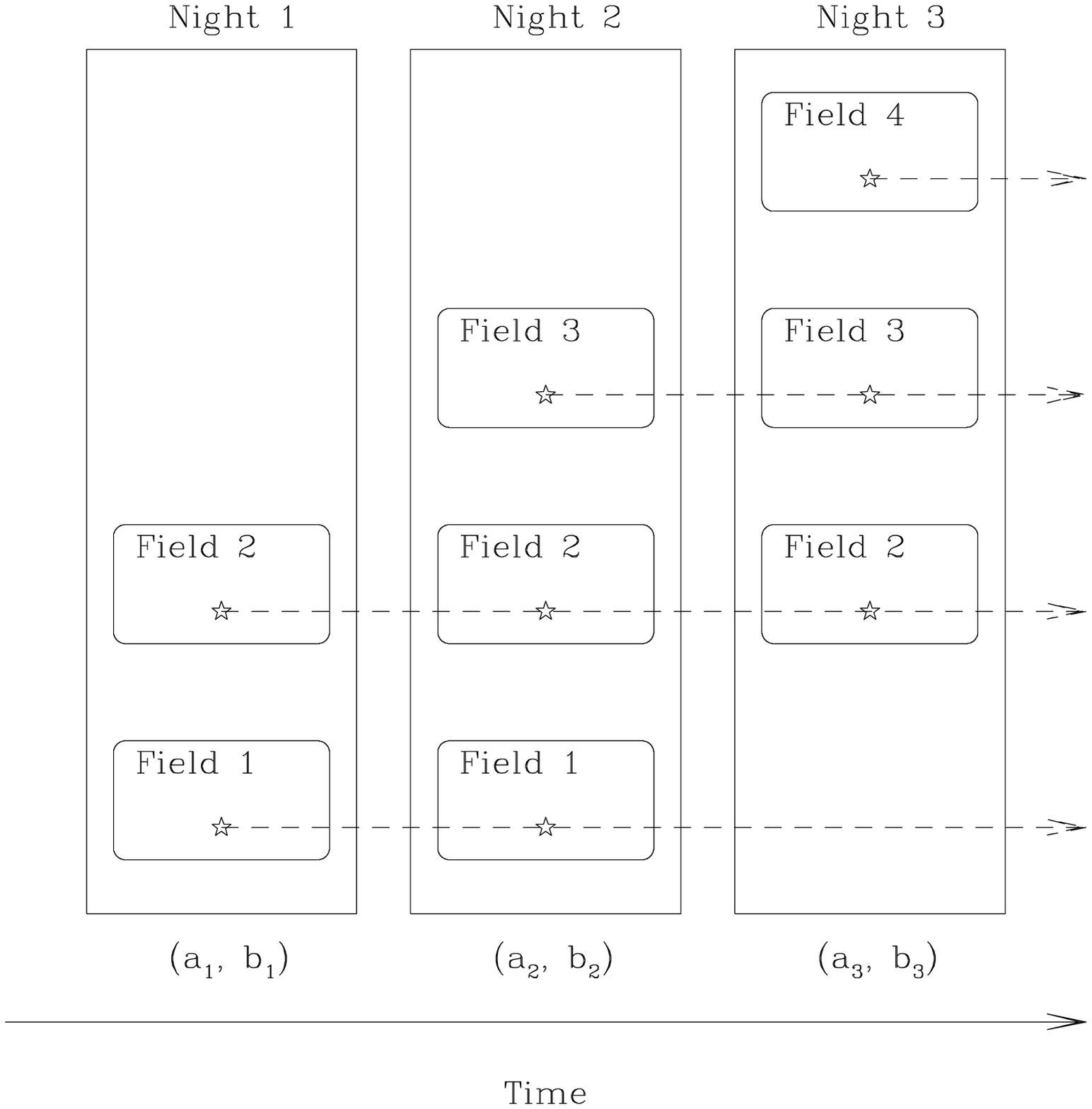}}
\caption{The essence of global calibration: photometry is bootstrapped
for all fields, sources and nights (see text). \label{fig1}}
\end{figure}

\begin{figure}
\epsfysize=8.5cm
\centerline{\epsfbox{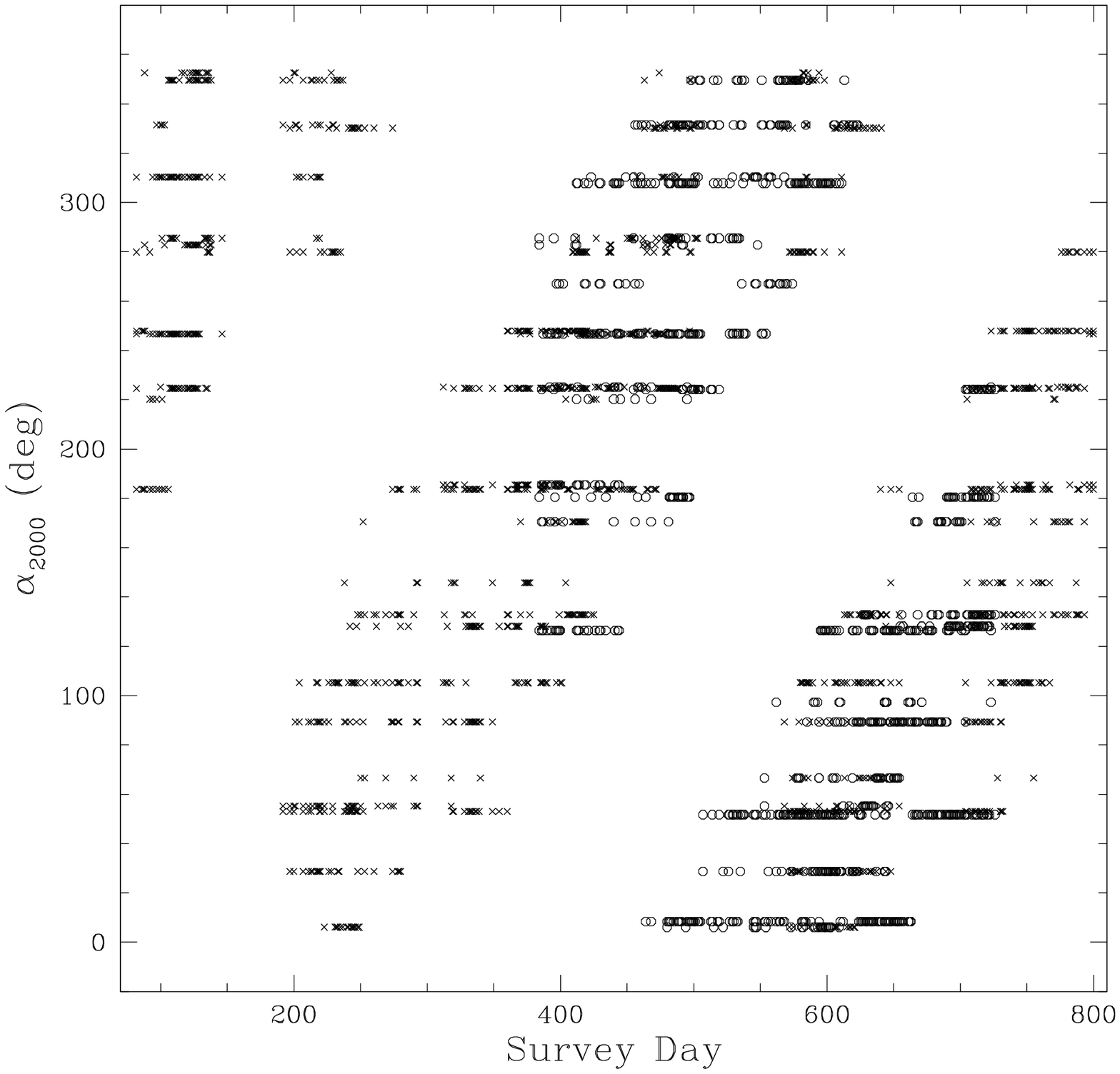}}
\caption{2MASS calibration coverage as the function of survey day (survey 
day `0' is defined to be March 1, 1997 UT).  The R.A. coordinates of the 
calibration fields are shown as they are observed in the course of the 
survey (cf. Figure~\ref{fig1}).  Crosses and circles denote calibration 
fields observed by Northern and Southern 2MASS facilities, respectively. 
\label{fig2}}
\end{figure}

The global calibration procedure~(\ref{eq2}) provides a better photometric 
solution than nightly calibration, since 
it minimizes residuals globally.  The benefit of global calibration is conveyed 
by Figure~\ref{fig1}: each night's photometry is calibrated through nightly 
parameters $a$ and $b$, tying fields observed that night to each other, and 
different nights are tied together by global 
parameters $m^{cal}$ and $A$.  For comparison, Figure~\ref{fig2} shows the 
actual survey sky coverage of calibration fields as a function of time.  

In practice, the set of equations~(\ref{eq2}) is written for a group of 
primary standards and so-called `tracer' stars, selected among relatively
bright field stars.  The same number of tracers $\mu$ is used in each 
calibration field.  In principle, any number of tracers can be used to 
derive the global calibration solution.  However, the large data sets 
associated with the problem prohibit using more 
than a few tracers per calibration field.  The set of equations is solved 
by a standard linear least squares algorithm (see Appendix~\ref{sec:details}), 
producing a set of photometric nightly constants $a_n$, $b_n$ for all survey 
nights, values of atmospheric extinction $A$, and calibrated magnitudes 
$m^{cal}$ for primary standards and tracer stars.  Once photometry is 
established for the primary and tracer stars in the field, it can be directly
applied to any of the other stars in the field.


\section{Applications of GPC} \label{sec:applications}

This section lists several applications of the global photometric calibration:
establishment of secondary standards, study of atmospheric extinction and 
demonstration of temporal and spatial consistency.  
The discussion in this section is based on a 
global calibration solutions for Northern data taken from May 1997 to May 1999, 
and Southern data from March 1998 to February 1999.  Calibration solutions for 
North and South are calculated separately and compared through common 
calibration fields.  The global solution was obtained in two steps.  First, 
we obtained a preliminary calibration solution based on the entire set of 
nights.  This preliminary solution was used to create nightly calibration 
plots similar to the one shown in Figure~\ref{fig:calnight}.  The nightly
plots were examined and the few with maximal deviations $>0.05$ mag from 
the linear behavior were removed from the data set.  We then reran the procedure 
on the new `clean' data to obtain the final calibration solution.
Northern data consisted of a total of 14,833 calibration scans of 26 fields taken
on 298 nights.  Solving equation~(\ref{eq2}) by a linear least squares 
algorithm amounts to inverting a sparse $37758 \times 675$ matrix.  The 
dimensions of the matrix, $N\times M$, are the total number of observations
$N$ and the number of free parameters, $M=k(\mu+1) + 2n + 1$, where $k$ is the
number of calibration fields, $\mu$ is the number of tracer stars in each field,
and $n$ is the number of survey nights.  In the Southern data, there are 227 
observing nights for 26 fields (11945 total calibration scans) and the solution
is found by inverting a smaller, $33407 \times 533$ matrix.  Both hemispheric 
solutions are obtained based on the photometry of fiducial tracers and two 
tracer stars in each calibration field (i.e., $\mu=2$).
In solving equation~(\ref{eq2}), we look for an unconstrained solution (i.e.
calibrated magnitudes of primary standards are free parameters; see 
\S\ref{sec:details}), to ensure no bias enters from the  known magnitudes of 
primary 
standards.  Because the solution is unconstrained, the GPC calibrated 
magnitude system will have an arbitrary constant offset in each band.
The offset for each band is found from averaging the difference between
GPC solution and a priori catalog magnitudes of the fiducial standards in 
Table~\ref{table1}.  The GPC system magnitudes for the primary standards in
each field are given in Table~\ref{tab:ss}.


\subsection{Secondary Standards} \label{sec:secondaries}

\begin{landscape}
\begin{figure*}
\vskip5cm
\mbox{
\mbox{\epsfysize=9.8cm\epsfbox{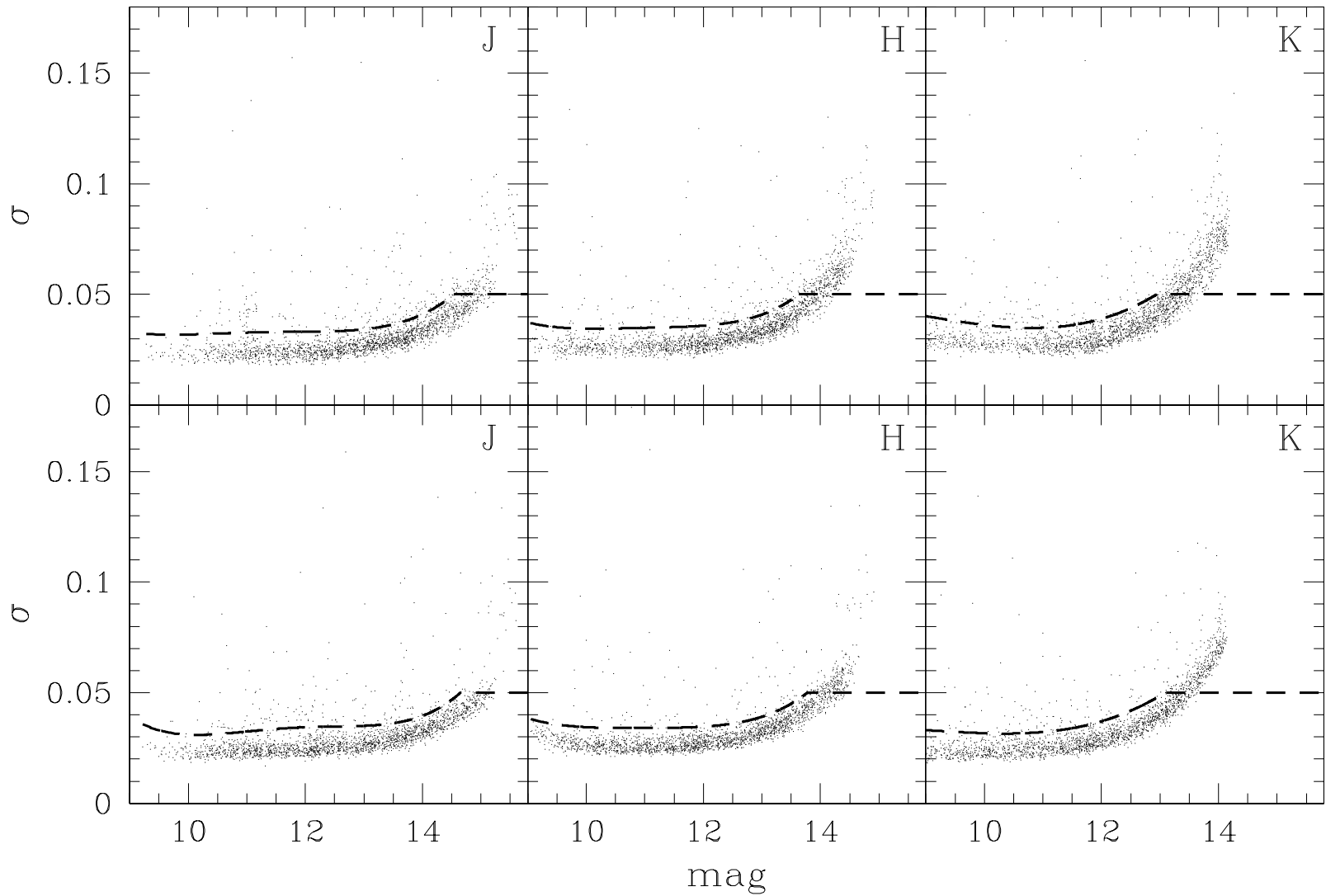}}
}
\caption{Pooled rms error plots for both hemispheric solutions. {\em Top row:}
Northern solution; {\em Bottom row:} Southern solution.  Each dot represents
one star.  Dashed lines show the adopted thresholds (see text).
\label{fig:pooled}}
\end{figure*}
\end{landscape}

To establish candidate secondary standards in the calibration fields, we apply
the GPC solution to additional field stars.  In order to exclude variable stars, 
secondary standard candidates must have a low rms uncertainty,
\begin{equation}
\sigma = \sqrt {\frac{1}{N} \sum _{i=1} ^N (m_i-\overline m)^2},
\end{equation}
where $N$ is the number of individual observations (scans) of a source.
To ensure the statistical significance, we keep only the candidates with
$N > 100$.  Figure~\ref{fig:pooled} shows the pooled rms plots for all 
solutions.  Each point in the plot represents a single star.  Dashed lines,
which indicate the adopted thresholds for rms error $\sigma$, were obtained
by fitting a fourth-order polynomial\footnote{A third-order polynomial was 
used in $K_s$ band} $f (m)$ to the data and adjusting its intercept point so 
that the number of points below the curve is $66\%$ ($1 \sigma$) of the total.
Toward fainter magnitudes $f (m)$ rises sharply, so we modify our 
$\sigma$-threshold to pick sources with
\[
\sigma < \min {\{f (m), 0.05\}}.
\]
The above selection criteria
produce a sample of low-variance stars, which are good candidates 
for secondary standards.  The high-variance sources that fail our 
$\sigma$-criterion are often variable stars, which can be reliably 
identified by examining their light curves.

After applying the GPC solution to each hemisphere, we obtain two lists
of candidate secondaries, one for each hemisphere.  The magnitudes of stars
in equatorial fields observed from both hemispheres are merged using
variance weighted average:
\begin{equation}
m = \frac {\left(\frac{m_N}{\sigma_N^2}+\frac{m_S}{\sigma_S^2}\right)}
{\left(\frac{1}{\sigma_N^2}+\frac{1}{\sigma_S^2}\right)},
\end{equation}
where $m_N$ and $m_S$ are average magnitudes for the star in the Northern and
Southern solutions, and $\sigma_N$, $\sigma_S$ are respective root variances.
The list of 2177 candidate secondary standards for all calibration fields is 
given in Table~\ref{tab:ss} in Appendix~\ref{sec:secstd}.  This list is used 
by the 2MASS Project to select a subset of unconfused point sources for actual 
calibration of 2MASS data.


\subsection{Atmospheric Extinction Coefficients} \label{sec:extinction}

\def\co2{\mbox{CO$_2$}}
\def\h2o{\mbox{H$_2$O}}

The GPC solution provides an estimate of the atmospheric extinction in all
three bands.  The atmospheric extinction in the near-infrared is caused 
primarily by molecular bands of \h2o (Hall \& Genet 1982, Mountain \etal{} 
1985) and, to a lesser extent, by \co2 (Manduca \& Bell 1979).  
In equation~(\ref{eq2}) extinction enters through the
coefficient $A$, which represents the average value of the atmospheric
extinction over the period sampled (one number for each band).  To 
decouple the 

\noindent atmospheric extinction parameters from photometric zero
points (sensitivities), the observations during each night span a range
of air masses ($X$).  A more detailed model for atmospheric extinction, 
for example involving a periodic function to mimic seasonal variations, 
could also be considered in equation~(\ref{eq2}).  However, any extinction 
model can be devised and applied a posteriori, once we compute the residuals 
between the regressed data and the observations, 
$\Delta m = m_{cal} - m_{inst}$.  One such model is presented in 
\S\ref{sec:uniformity}.  

\begin{figure}[h!]
\mbox{
\mbox{\epsfysize=8.3cm\epsfbox{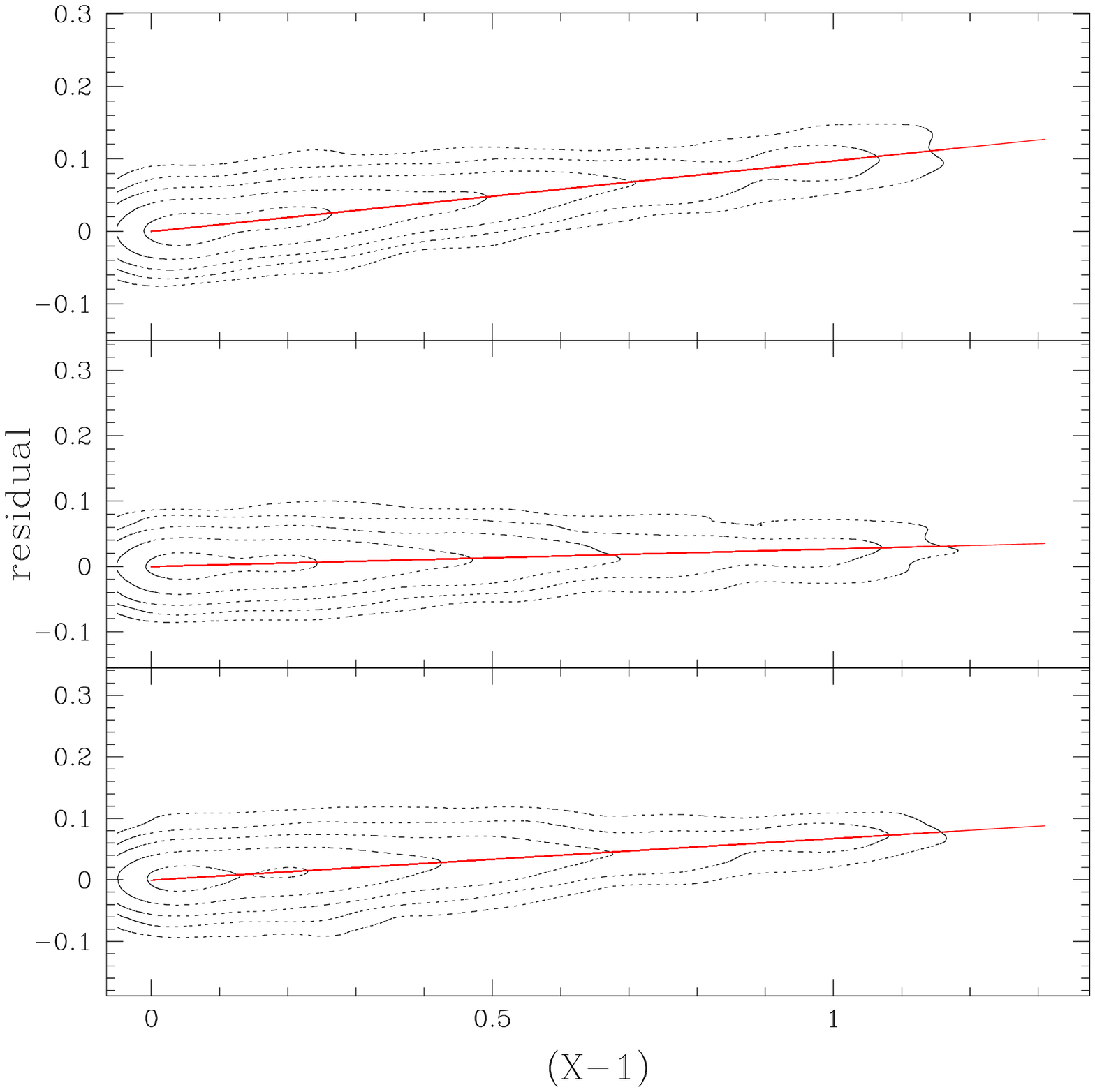}}
\mbox{\epsfysize=8.3cm\epsfbox{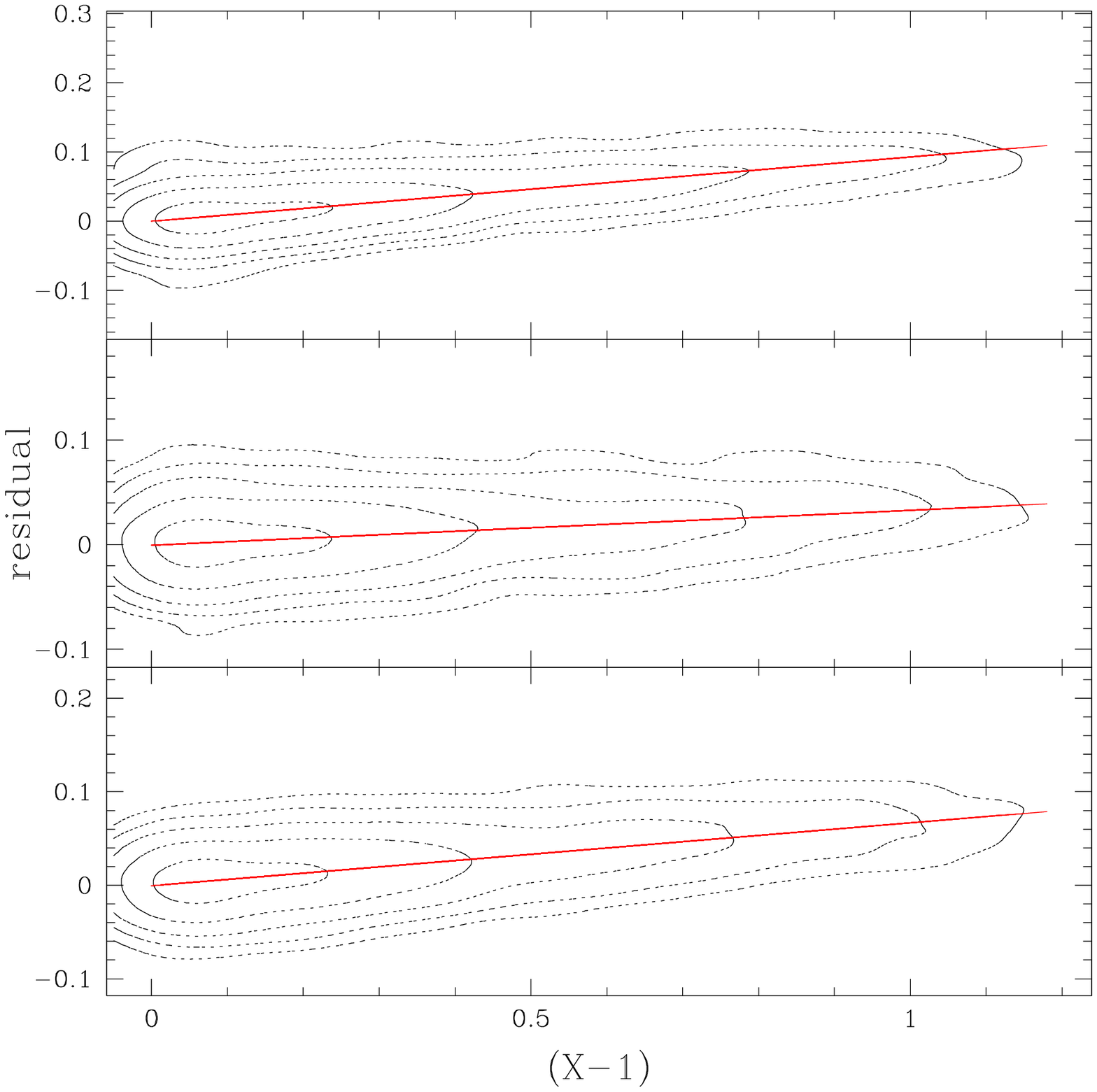}}
}
\caption{Density plots of the residuals (without the extinction term)
as a function of the airmass. {\em Left:} Northern solution; {\em Right:} 
Southern solution.  Vertical axis is in units of magnitudes.  Contour 
levels are logarithmic, corresponding to point density of 0.3, 1, 3, 10 and 
30 mag$^{-1}$ airmass$^{-1}$. The slopes of the straight lines are the 
derived mean atmospheric extinction coefficients listed in 
Table~\ref{table:ext}.  \label{fig:atmext}}
\end{figure}

The atmospheric extinction coefficients for North and South produced 
by GPC are given in Table~\ref{table:ext}, which also lists CTIO values 
from Frogel~(1995).  The uncertainties in extinction coefficients are 
random errors, obtained by inversion of the Hessian matrix.
\begin{table}[b!]
\centering
\caption{Atmospheric extinction coefficients and mean 
squared norms of the residual vectors from GPC.  The extinction coefficients
are in magnitudes per unit airmass.  The CTIO values from Frogel (1998) are
listed for comparison. \label{table:ext}}
\begin{tabular}{cccc} \hline\hline
Band  & Hemisphere & 2MASS $A_\lambda$ & Frogel $A_\lambda$, CTIO \\ \hline
$J$ & North & $0.096\pm0.001$ & \\
$J$ & South & $0.092\pm0.001$ & $0.100\pm0.024$ \\
$H$ & North & $0.026\pm0.001$ & \\
$H$ & South & $0.031\pm0.001$ & $0.055\pm0.021$ \\
$K_s$ & North & $0.066\pm0.001$ & \\
$K_s$ & South & $0.065\pm0.001$ & $0.085\pm0.018$ \\ \hline
\end{tabular}
\end{table}
The values listed in the table are consistent at the $2-\sigma$ level.  
Figure~\ref{fig:atmext} shows the residuals $m_{inst}-m_{cal}-a-b \Delta t$
as a function of airmass.  The slopes of the linear regression lines are
the mean extinction coefficients.

\begin{figure}[thb]
\mbox{
\mbox{\epsfysize=8.3cm\epsfbox{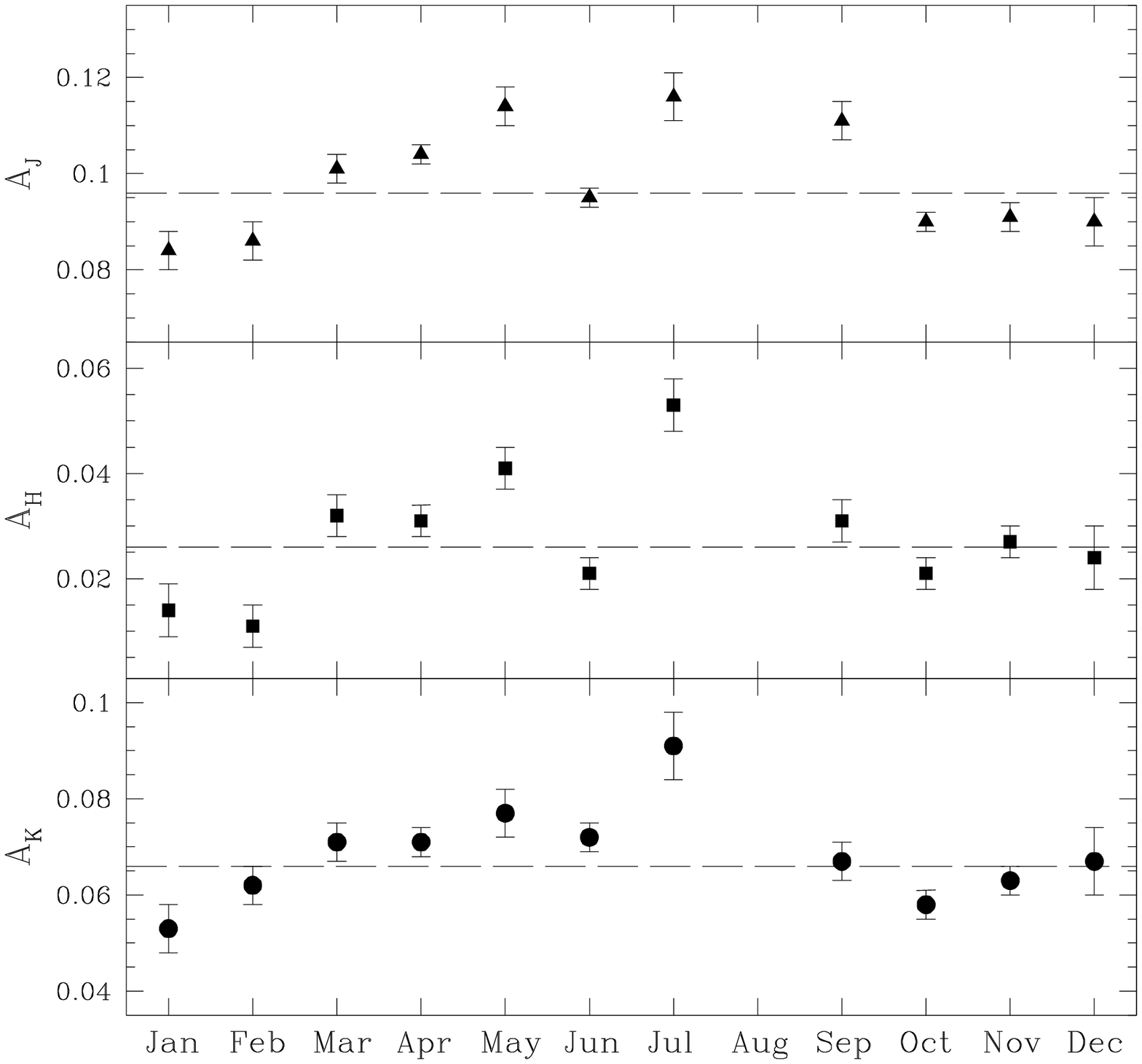}}
\mbox{\epsfysize=8.3cm\epsfbox{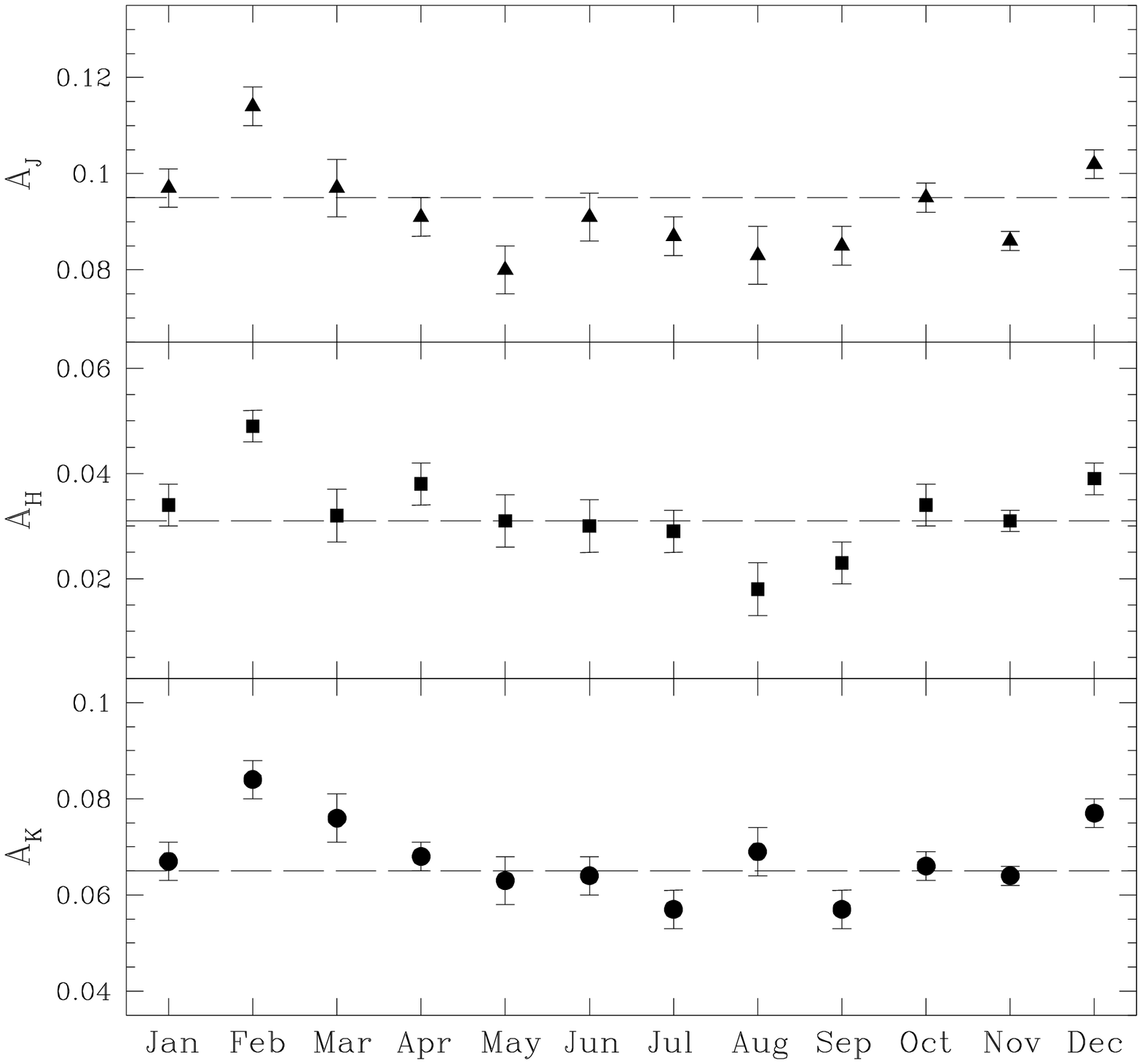}}
}
\caption{Seasonal variations in the atmospheric extinction (in mag/airmass)
in the North
({\em left panel}) and in the South ({\em right panel}).  Monthly average 
coefficients in $J$, $H$, and $K_s$ are plotted with triangles, squares, and 
circles, respectively.  The errors are indicated.  Horizontal dashed lines 
show global averages in three bands derived from the respective GPC 
solutions.  No August data for the Northern solution were available.
\label{fig:monthly}}
\end{figure}

Given the time baseline of 2MASS and the homogeneity and accuracy of 2MASS
photometric data, we have a great opportunity for studying atmospheric
extinction in the near-infrared over the period of more than a year.
Figure~\ref{fig:monthly} shows the seasonal variations in atmospheric
extinction.  The monthly averages were derived by, first, calculating the
residuals without the extinction term for each month separately, and then, 
by fitting a straight line to the distribution of residuals as a function 
of the airmass.  
While some scatter is present in Figure~\ref{fig:monthly}, the overall 
behavior of monthly extinction averages is in agreement with 
expectations.  The water vapor content in the atmosphere peaks during 
Northern spring and summer, which leads to higher atmospheric extinction 
in the North during that season.  The amplitude of the seasonal changes 
is about $0.02$ magnitudes/airmass.


\subsection{Global Consistency} \label{sec:uniformity}

The global consistency of the solution can be assessed from the analysis of residuals
as a function of spatial coordinate (spatial uniformity) or time (temporal
uniformity).  In this section, we present the results of two such analyses.
First, we examine the uniformity of the standards between the hemispheres, from looking 
at residuals $\Delta m = m_{North} - m_{South}$ for common calibration fields.
Then, we test spatial and temporal uniformity by comparing the calibrated
magnitudes of primary standards (derived from GPC) with literature photometry.

\subsubsection{North vs. South Uniformity}

\begin{figure}[h!]
\epsfysize=8.3cm
\centerline{\epsfbox{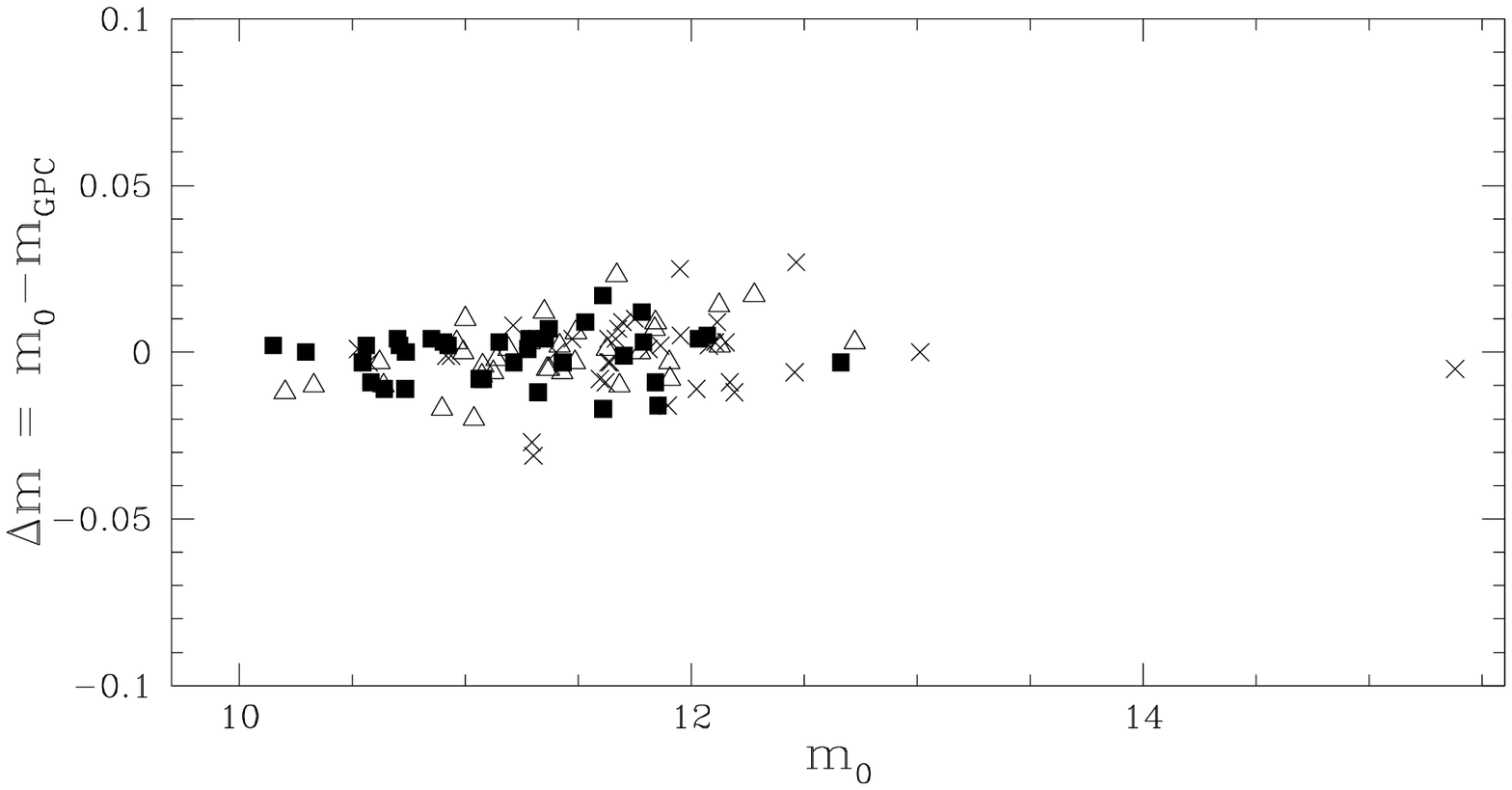}}
\caption{The difference between catalog and GPC magnitudes as a function of
catalog magnitude for primary standards.  Symbols represent $J$, $H$, and $K_s$
magnitudes (crosses, triangles and squares, respectively).  Most data points 
differ from zero by less than 0.02 mag.  Note the absence of any trends over
the entire magnitude range.
\label{fig:catgpc}}
\end{figure}

The linearity of the GPC solution is demonstrated in Figure~\ref{fig:catgpc},
which shows the difference between catalog and GPC magnitudes for primary 
standards as a function of the catalog magnitude.
The figure shows that the absolute value of the difference for most primaries
is less than 0.02 magnitude, i.e. consistent with zero.  The absence of any
trends in these data demonstrates the linearity of the GPC solution.  Note that 
the results of such comparison are limited by the accuracy of input catalog
since the errors in the fiducial magnitudes of primary standards are propagated
unchanged.  In fact, the good agreement demonstrates the high accuracy of the 
input catalog.

\begin{figure}[thb]
\mbox{
\mbox{\epsfysize=8.3cm\epsfbox{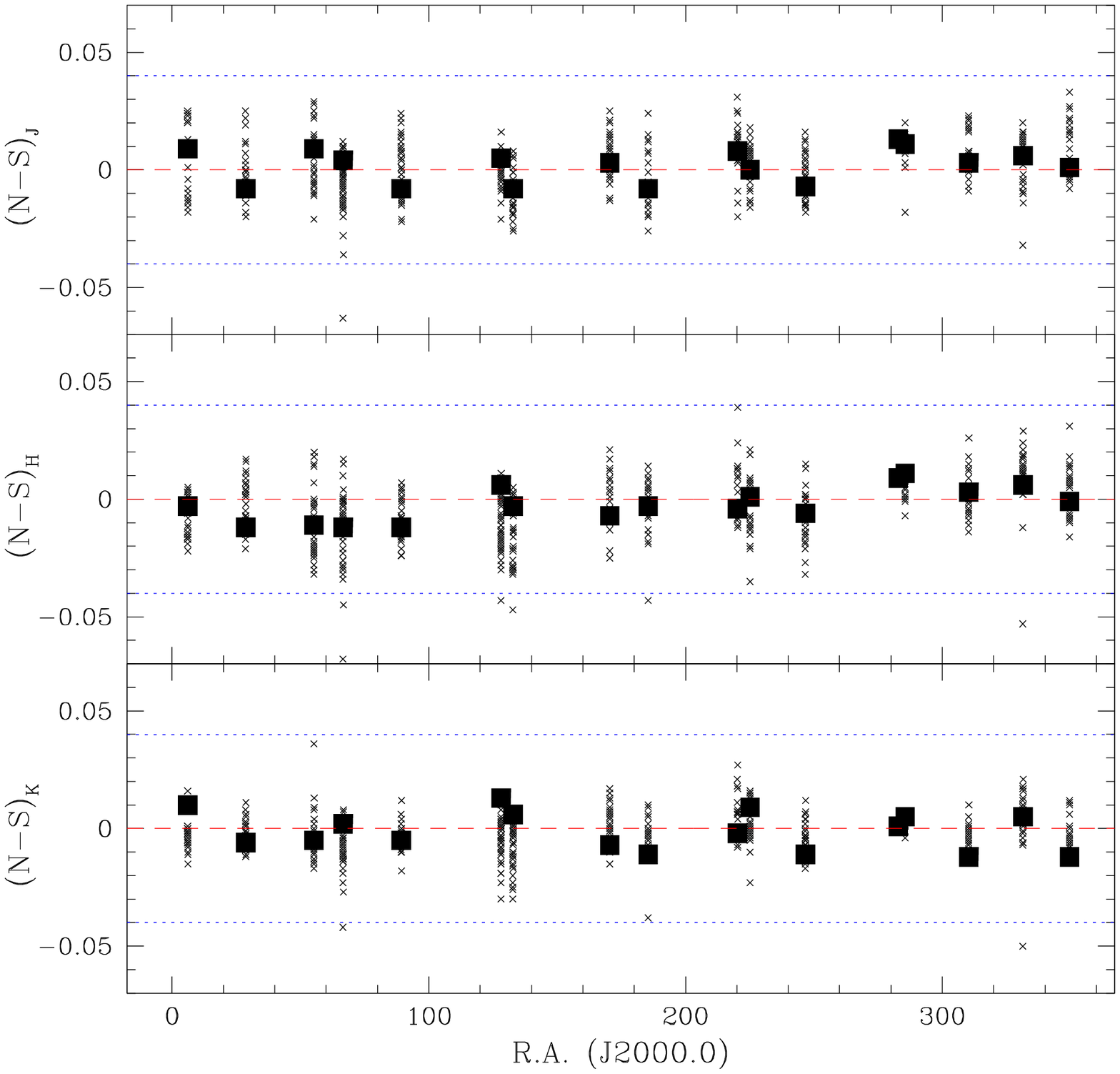}}
\mbox{\epsfysize=8.3cm\epsfbox{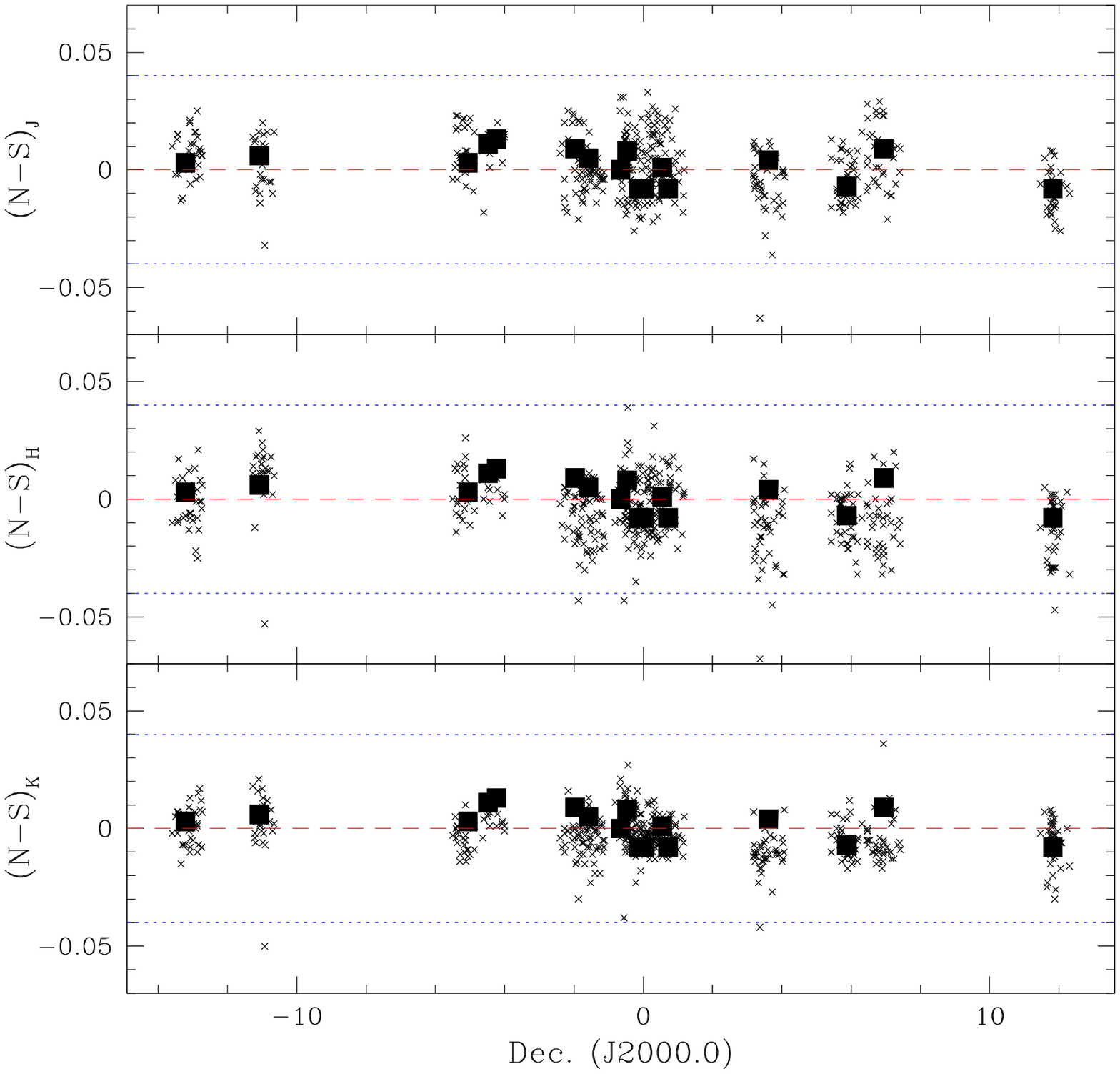}}
}
\caption{`North-South' difference between calibrated magnitudes of 
stars in common fields as a function of right ascension ({\em left panel})
and declination ({\em right panel}).  Crosses represent field stars, squares
represent primary standards.  Clusters of points are calibration fields.  
Due to elongated shape of calibration fields in latitude, the clusters are 
more dispersed in the right panel.  Dotted horizontal lines are drawn at 
$\Delta_m = \pm 0.04$, dashed line is at zero. \label{fig:ns_coord}}
\end{figure}

\begin{figure}
\epsfysize=12.5cm
\centerline{\epsfbox{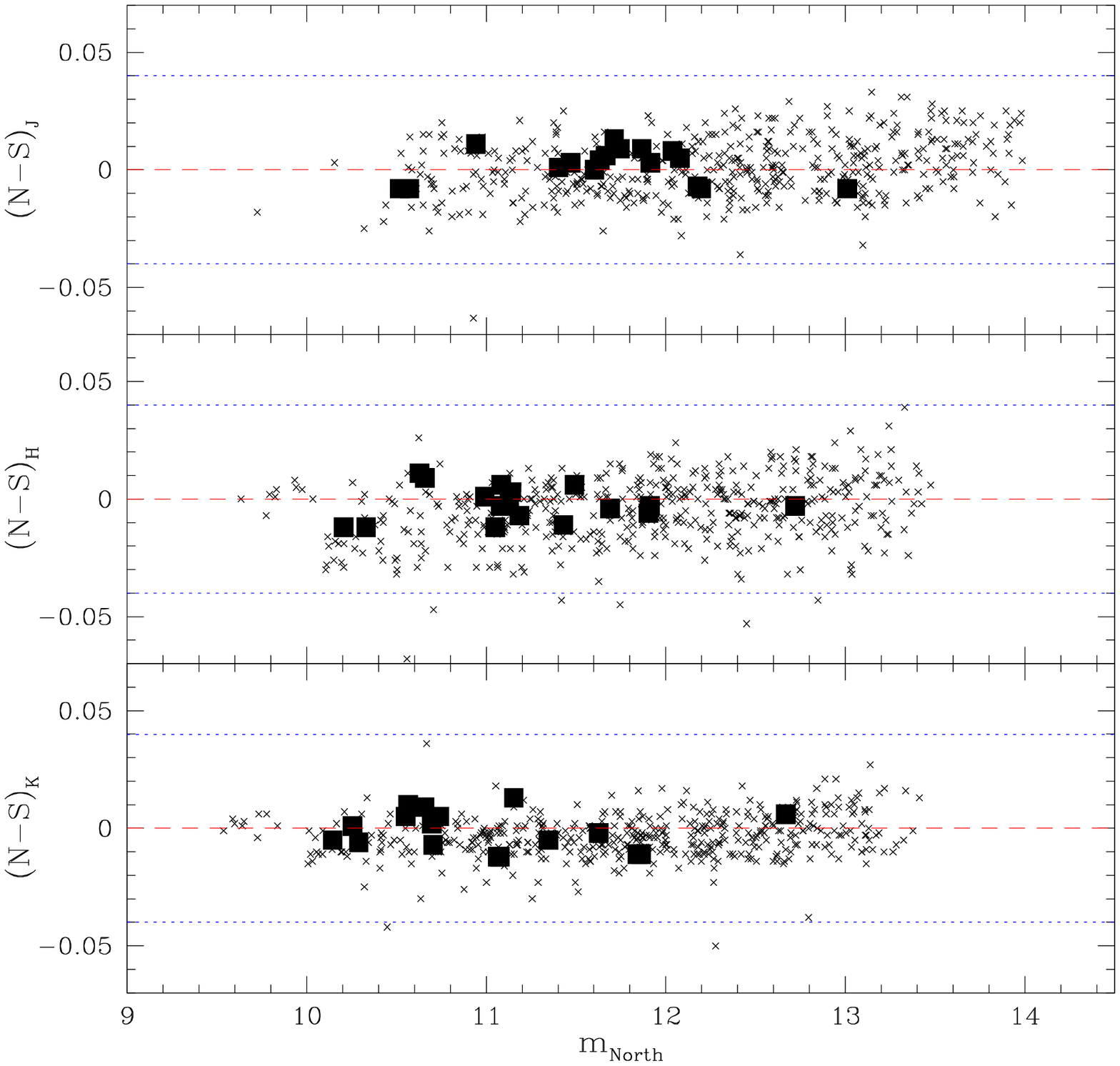}}   
\caption{Difference (in magnitudes) between Northern and Southern solutions 
as a function of Northern magnitude.  The notations are the same as in
Figure~\ref{fig:ns_coord}. \label{fig:ns_mag}}
\end{figure}

The comparison of the photometry between the hemispheres is carried out in 
equatorial fields (see Table~\ref{table1}).  Data from each 
hemisphere were calibrated separately by unconstrained GPC.  The solutions
(nightly photometric constants $a$ and $b$, and extinction parameter $A$)
were applied to the 100 brightest stars in each calibration field.
Figure~\ref{fig:ns_coord} shows the difference between calibrated 
magnitudes of common stars as the function of coordinate.  Except for a 
few outliers, the largest excursions for individual secondary standards 
(crosses) are bounded by $\Delta m < 0.04$, while the largest excursions for 
the primary standards (squares) are less than $0.02$ mag.
The mean differences for the photometry of common standards (squares in 
Figure~\ref{fig:ns_coord}) are $0.002\pm0.007$ ($J$), $-0.002\pm0.007$ 
($H$), and $-0.001\pm0.008$ ($K_s$).  The plot of North-South photometric 
offset vs. declination displays a slight trend of unknown origin.  The 
slopes of the corresponding regression lines in $J$, $H$, and $K_s$ are 
$(-4.9\pm0.9)\times10^{-4}$ mag/deg, $(-7.5\pm0.9)\times10^{-4}$ mag/deg, 
and $(-3.9\pm0.6) \times10^{-4}$ mag/deg, respectively.  The same regression 
analysis using primary
standards only gives slopes consistent with zero at the $2\sigma$ level: the 
corresponding slopes are $(-4.9\pm2.8)\times10^{-4}$ in $J$, 
$(-5.2\pm2.8)\times10^{-4}$ in $H$, and $(-0.4\pm3.5)\times10^{-4}$ in $K_s$.
Figure~\ref{fig:ns_mag} shows the same residuals as in 
Figure~\ref{fig:ns_coord}, except as a function of magnitude in the North.
The slopes of the regression lines in all three bands are consistent with
zero.

\subsubsection{Temporal Stability} \label{sec:drift}

\begin{figure}[h!]
\mbox{
\mbox{\epsfysize=8.3cm\epsfbox{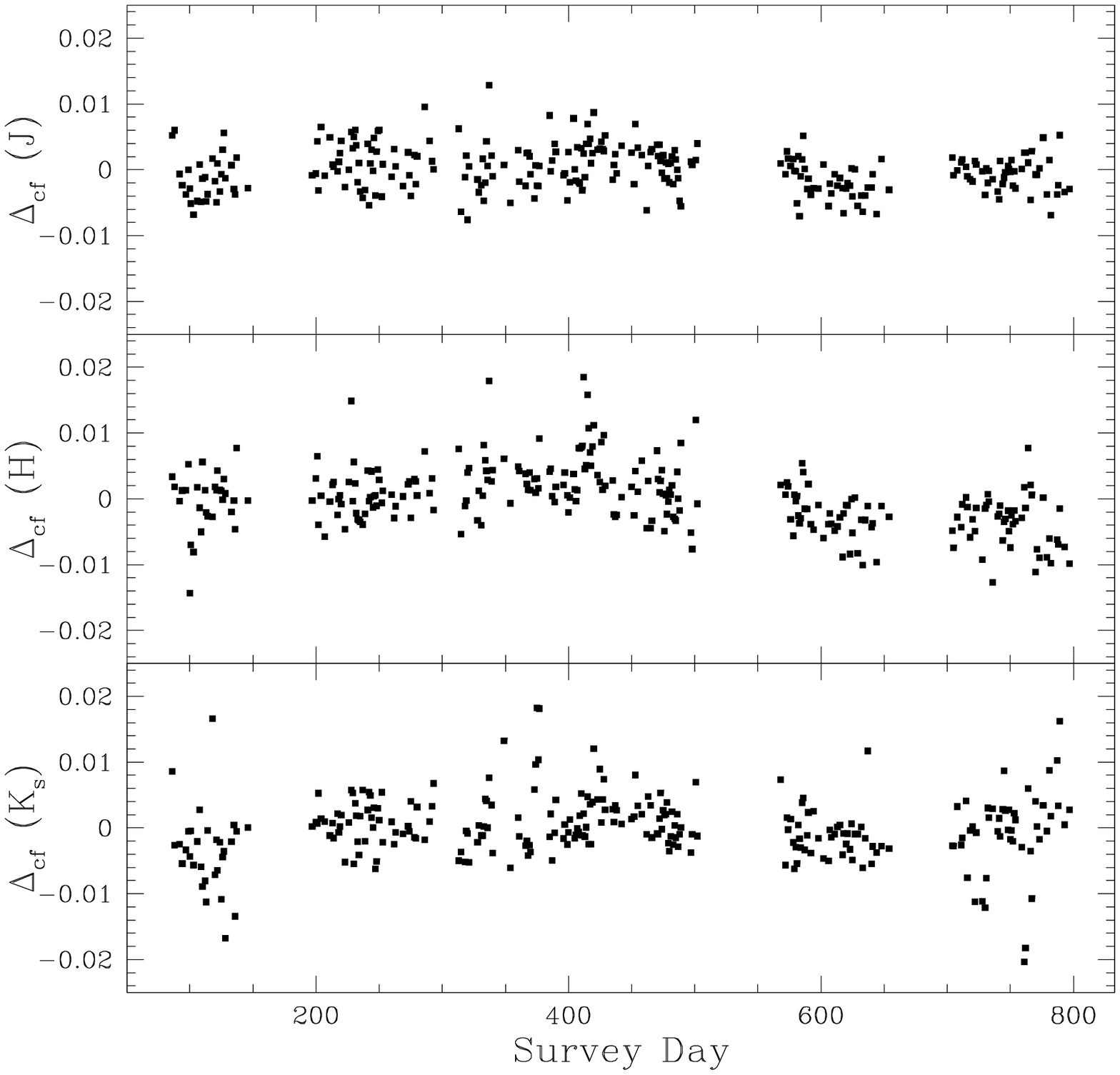}}
\mbox{\epsfysize=8.3cm\epsfbox{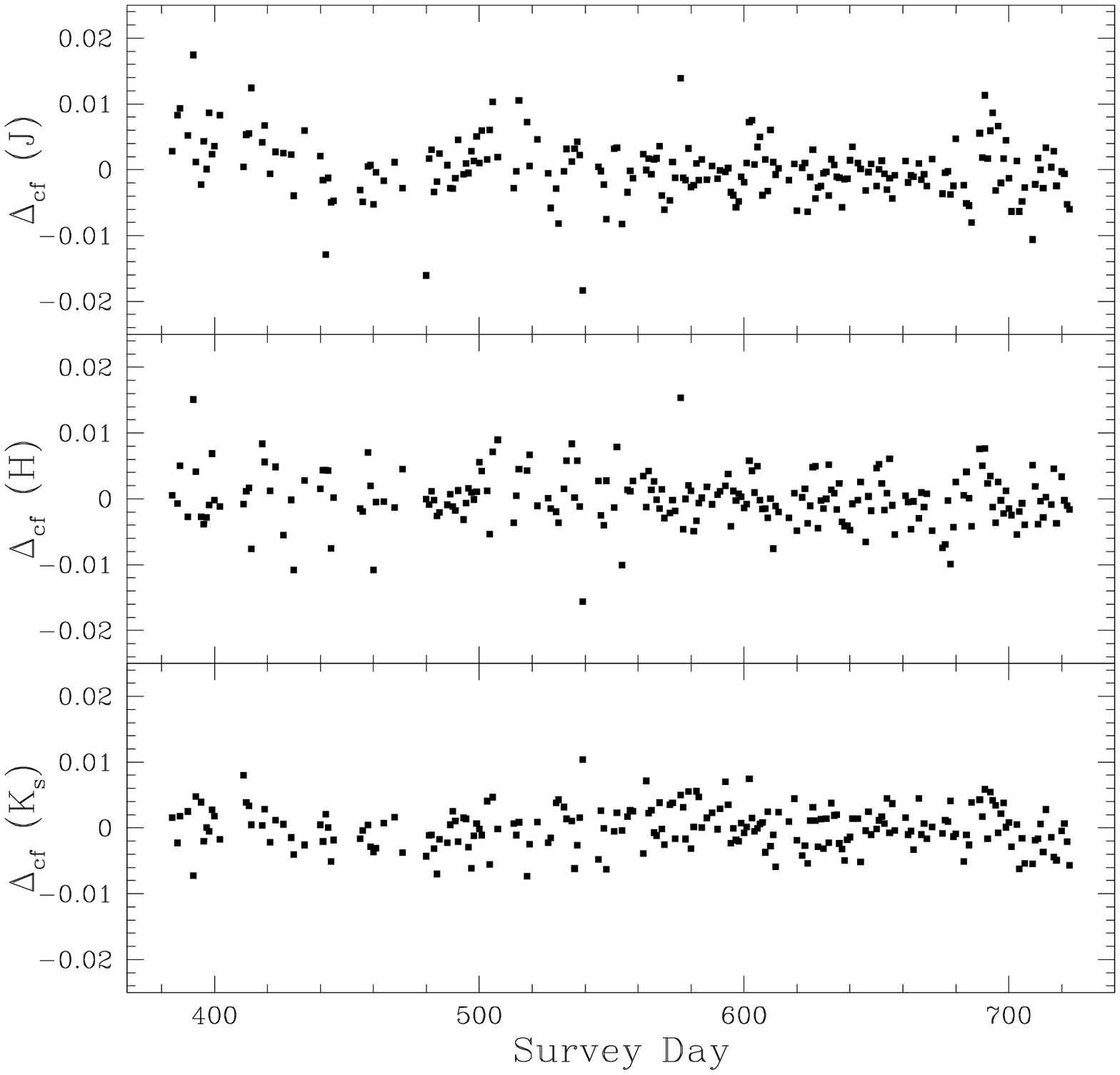}}
}
\caption{Difference between calibrated and fiducial magnitude for
primary standards as a function of time for Northern ({\em left panel})
and Southern ({\em right panel}) solutions.  Each survey night is 
represented by a single point which displays the average difference 
for all standards observed on that night.
\label{fig:cf_time}}
\end{figure}

To analyze the temporal stability, we consider the difference between 
calibrated and fiducial magnitudes of primary standards as a function of
time.  Specifically, we calculate the difference `calibrated-fiducial' for
all primary standards observed on a given night and plot the average
difference.  Figure~\ref{fig:cf_time} shows the corresponding differences 
in all three bands for Northern and Southern hemispheres.
The figure, which essentially indicates the long-term stability of the global 
calibration, shows a very small temporal variation in the photometry
of the primary standards.  The amplitude of the variations is less than
$1\%$.


\section{Summary} \label{sec:summary}

\begin{itemize}
\item  We have presented a global photometric calibration procedure and 
applied it to more than a year of 2MASS calibration data.  Exploiting 
the fact that the observed nightly zeropoint drift is linear, one can
simultaneously calibrate the photometry of fiducial standards and any 
number of tracer stars.  The solution is found by using a standard 
least squares 
algorithm and is {\em globally} optimized, in the sense that it is 
best solution for {\em all} survey fields and nights.  The solution 
does not presume a priori knowledge of the magnitudes for any of the 
stars, but instead produces the optimal set of relative magnitudes.
A constant offset is determined by averaging the differences between
the calibrated and catalog magnitudes for the primary standards.

\item  The photometry of the primary and two tracer stars per field 
is transferred to the 100 brightest non-variable stars in each field.
With a $\sigma$-threshold, we select a low-dispersion sample of 2177 
candidate secondary standards to accompany the set of primary 2MASS 
standards.  The selected secondaries fall in the magnitude range 
$9.0<J<14.5$, $9.0<H<13.9$, $9.0<K_s<13.7$.

\item Using the global solution, we derive the atmospheric extinction.
The global averages are $A_J/A_H/A_{K_s}=0.096/0.026/0.066$ (North), and
$A_J/A_H/A_{K_s}=0.092/0.031/0.065$ (South).  The averages are consistent
with previous work at the $2\sigma$ level.

\item Seasonal variations in the atmospheric extinction are analyzed.
The amplitude of the variations is approximately $0.02$ mag/airmass
and is consistent with the behavior expected from variations of the water
vapor content in the atmosphere.
\end{itemize}

\section*{Acknowledgements}

The authors would like to thank the IPAC staff for preparing the tapes 
with calibration data and for their assistance with the study.
SN and MDW acknowledge funding through a NASA/JPL grant to 2MASS Core
Project science.
RMC, SLW and JEG acknowledge support from the Jet Propulsion Laboratory,
which is operated under contract from NASA by the California Institute of
Technology.  This publication makes use of data products from the Two Micron 
All Sky Survey, which is a joint project of the University of Massachusetts 
and the Infrared Processing and Analysis Center, funded by the 
National Aeronautics and Space Administration and the National 
Science Foundation.


\begin{appendix}
\section{Least Squares} \label{sec:details}
The set of equations~(\ref{eq2}) represents a linear least squares (LLS) 
problem.  The general LLS problem is well-documented in the literature (e.g., 
Lawson \& Hanson 1973, Greenbaum 1986) and has many professional codes 
available for its solution.  Written in the matrix form, the mathematical 
model for the GPC is
\begin{equation}
\Phi {\bf x} = {\bf b}, \label{eq:matrix}
\end{equation}
where $\Phi$ is the design matrix, ${\bf x}$ is the parameter vector and 
${\bf b}$ is the vector of observed (instrumental) magnitudes.  The 
parameter vector ${\bf x}$ consists of the magnitudes of the fiducial 
standards in each field, followed by the magnitudes of the tracer stars 
for each field, followed by the photometric nightly constants $a$, $b$ for 
each night and the atmospheric extinction coefficient $A$:
\begin{equation}
{\bf x}^T = \left(
m^{std}_1, \ldots, m^{std}_k,
m^0_{11}, \ldots, m^0_{\mu 1}, \ldots, m^0_{1 k}, \ldots, m^0_{\mu k},
a_1, \ldots, a_n,
b_1, \ldots, b_n,
A \right), \label{eq:parvect}
\end{equation}
where $k$ denotes the total number of calibration fields and $n$ denotes the 
total number of observation nights.  The number of tracers in the field is $\mu$ 
(the same number of tracers in each field is used).  The design matrix $\Phi$ 
has the following structure:
\begin{displaymath}
\Phi = \left(
\begin{array}{ccccc}
\overbrace{\mathstrut 1, 0, \ldots, 0}^{k\ \mathrm{standards}} &
\overbrace{\mathstrut 0, 0, \ldots, 0}^{\mu \times k\ \mathrm{tracers}} &
\overbrace{\mathstrut 1, 0, \ldots, 0}^{n\ \mathrm{nights}} &
\overbrace{\mathstrut \Delta t_{jn}, 0, \ldots, 0}^{n\ \mathrm{nights}} &
X_{jn}-1 \\
{0, 1, \ldots, 0} & {0, 0, \ldots, 0} & {1, 0, \ldots, 0} & 
{\Delta t_{jn}, 0, \ldots, 0} & X_{jn}-1 \\
{0, 1, \ldots, 0} & {0, 0, \ldots, 0} & {0, 1, \ldots, 0} & 
{0, \Delta t_{jn}, \ldots, 0} & X_{jn}-1 \\
{\ldots} & {\ldots} & {\ldots} & {\ldots} & {\ldots} \\
{0, 0, \ldots, 0} & {1, 0, \ldots, 0} & {1, 0, \ldots, 0} & 
{\Delta t_{jn}, 0, \ldots, 0} & X_{jn}-1 \\
{\ldots} & {\ldots} & {\ldots} & {\ldots} & {\ldots}.
\end{array}
\right)
\end{displaymath}
The tracer stars are grouped by the fields, i.e. after the fiducial 
standards, there are $n$ tracers from the first field, then there are $n$ 
tracers from the second field, and so on.  There is a single unity in the 
first $k (\mu+1)$ columns of each row of the design matrix, representing
the observation of either the fiducial standard or a tracer star. The 
subscript $j$ denotes the time offset from the midnight on the night $n$.
The dimensions of the matrix are $N \times M$, where $N$ is the total 
number of observations of all fiducial standards and tracers, and 
$M = k (\mu+1) + 2n + 1$ is the number of free parameters.

\subsection{Constrained Scheme}
The least squares solution of the equations~(\ref{eq2}) is globally optimal 
(i.e., it minimizes rms), but, generally speaking, has an arbitrary offset 
from the true magnitude:
\begin{equation}
\left(m^{\lambda, 0}_{ik}\right)_{true} = 
\left(m^{\lambda, 0}_{ik}\right)_{sol} + K_{ik}. \label{eq:offset}
\end{equation}
To produce the optimal solution without the arbitrary additive constant 
$K_{ik}$ one has to set the magnitudes of the fiducial standards as fixed 
points of the GPC.  This results into GPC with constraints, where 
the constraint equations are:
\begin{equation}
m^{std}_{ijkn} = m^{std, 0}_{ik} \label{eq:cons}
\end{equation}
(the superscript $std$ denotes the fiducial standards in the fields). In the 
matrix form, the calibration problem with constraints is written as
\begin{eqnarray}
\Phi {\bf x} = {\bf b}, \nonumber \\
C {\bf x} = {\bf d}. \nonumber
\end{eqnarray}
For this particular problem, the constraint equations~(\ref{eq:cons}) result 
in the identity constraint matrix $C$, and the LLS problem with constraints 
is straightforward to solve using, e.g., orthogonal basis for the null space 
of the constraint matrix (Lawson \& Hanson 1973, Ch. 20).  The GPC with fixed 
points derives the globally optimal solution which is the closest to the
listed magnitudes of the fiducial standards.

\section{Secondary Standards} \label{sec:secstd}

Below, we present a sample table of 2MASS secondary standards, which 
includes the primary and some of the secondaries for the field 90021.  
The primary standard is the first star listed in each field in 
Table~\ref{tab:ss}.  Columns in the 
table are $\alpha_{J2000}$, $\delta_{J2000}$ coordinates of the stars, 
the globally calibrated $JHK_s$ magnitudes with the corresponding 
uncertainties, and the number of observations (i.e. individual scans) 
in each band.  The positions of the secondaries are accurate with
respect to the ICRS to $<0.5''$ rms.  Finding charts for the secondaries
can be obtained by using 2MASS Visualizer tool at
{\tt http://irsatest.ipac.caltech.edu:8001/applications/2MASS/ReleaseVis/}\footnote{Finding charts are available only for calibration fields which 
are covered by the Second Incremental Data Release}.  The full table of 
candidate secondary standards in all calibration fields is available from 
FTP archives at {\tt ftp://nova.astro.umass.edu/pub/nikolaev/}, or at 
{\tt ftp://anon-ftp.ipac.caltech.edu/pub/2mass/globalcal/}.

\begin{table*}[h!]
\caption{2MASS Secondary Standards (sample).  Columns list J2000.0 coordinates
of the stars. $JHK_s$ magnitudes and their rms errors and the number of
observations in each band. \label{tab:ss}}
\begin{tabular}{rrccccccccc}
\hline\hline
$\alpha$ (J2000) & $\delta$ (J2000) & J & $\sigma_J$ & H &
$\sigma_H$ & $K_s$ & $\sigma_{K_s}$ & $N_J$ & $N_H$ & $N_{K_s}$ \\ \hline

\multicolumn{11}{c}{Field: 90021} \\
  6.10250 & $-1.97230$& 11.862& 0.016& 11.081& 0.016& 10.559& 0.018 &  793 &  823 &  817 \\
  6.03858 & $-2.26631$& 13.862& 0.029& 13.258& 0.032& 13.148& 0.044 &  450 &  464 &  463 \\
  6.04265 & $-2.28167$& 11.922& 0.026& 11.556& 0.025& 11.485& 0.026 &  740 &  767 &  762 \\
  6.04303 & $-2.02408$& 11.893& 0.019& 11.243& 0.019& 11.089& 0.019 &  748 &  775 &  770 \\
  6.04514 & $-1.71895$& 11.367& 0.019& 10.881& 0.019& 10.798& 0.019 &  768 &  798 &  792 \\
  6.05158 & $-2.04839$& 13.702& 0.022& 13.211& 0.023& 13.136& 0.030 &  781 &  811 &  805 \\
  6.05168 & $-1.86255$& 13.552& 0.022& 13.241& 0.024& 13.189& 0.033 &  781 &  811 &  805 \\
  6.05223 & $-2.01792$& 13.260& 0.019& 12.845& 0.021& 12.766& 0.025 &  781 &  811 &  805 \\
  6.05384 & $-2.27687$& 13.055& 0.025& 12.585& 0.029& 12.498& 0.032 &  785 &  815 &  809 \\
  6.05452 & $-1.83307$& 13.208& 0.021& 12.893& 0.024& 12.823& 0.028 &  787 &  817 &  811 \\
  6.06506 & $-2.16824$& 13.885& 0.025& 13.405& 0.027& 13.326& 0.033 &  793 &  823 &  817 \\
  6.07232 & $-1.62850$& 13.781& 0.022& 13.138& 0.025& 12.940& 0.027 &  793 &  823 &  817 \\
  6.07362 & $-2.16577$& 13.477& 0.023& 12.864& 0.022& 12.628& 0.025 &  793 &  823 &  817 \\
  6.07832 & $-1.69025$& 14.195& 0.026& 13.497& 0.029& 13.328& 0.032 &  793 &  823 &  817 \\
  6.08807 & $-2.31853$& 14.019& 0.031& 13.366& 0.035& 13.180& 0.041 &  793 &  823 &  817 \\
  6.08952 & $-1.87956$& 13.329& 0.020& 12.750& 0.021& 12.510& 0.024 &  793 &  823 &  817 \\
  6.09105 & $-2.34330$& 12.415& 0.023& 11.870& 0.025& 11.772& 0.026 &  793 &  823 &  817 \\
  6.09865 & $-1.55405$& 12.306& 0.018& 11.971& 0.022& 11.915& 0.022 &  793 &  823 &  817 \\
  6.09878 & $-1.98499$& 12.144& 0.016& 11.545& 0.017& 11.431& 0.018 &  793 &  823 &  817 \\
  6.10086 & $-2.39661$& 12.621& 0.020& 11.974& 0.020& 11.847& 0.021 &  793 &  823 &  817 \\
\hline
\end{tabular}
\end{table*}

\end{appendix}

\begin{thebibliography}{}
\bibitem[]{} Casali,~M. \& Hawarden,~T. 1992, JCMT-UKIRT Newsl., No. 4, 33
\bibitem[]{} Greenbaum,~A. 1986, {\em Iterative Methods for Solving Linear
Systems} (Philadelphia: SIAM)
\bibitem[]{} Hall,~D.~S. \& Genet,~R.~M. 1982, {\em Photoelectric Photometry
of Variable Stars} (Fairborn, OH: IAPPP)
\bibitem[]{} Lawson,~C.~L. \& Hanson,~R.~J. 1995, {\em Solving Least Squares
Problems} (Philadelphia: SIAM)
\bibitem[]{} Manduca,~A. \& Bell,~R.~A. 1979, PASP, 91, 848
\bibitem[]{} Mountain,~C.~M., Leggett,~S.~K., Selby,~M.~J., and Zadrozny,~A.
1985, A\&A, 150, 281
\bibitem[]{} Persson,~S.~E., Murphy,~D.~C., Krzeminski,~W., Roth,~M., 
Rieke,~M.~J. 1998, AJ, 116, 2475
\end{thebibliography}
\end{document}